\documentclass[submission,copyright,creativecommons]{eptcs}
\providecommand{\event}{NCMA 2023} 

\usepackage[T1]{fontenc}
\usepackage[english]{babel}
\usepackage{mathrsfs}
\usepackage{xcolor}
\usepackage{upgreek}
\usepackage{enumerate}
\usepackage{enumitem}
\usepackage{scalerel}
\usepackage{amssymb}
\usepackage{amsmath,amsthm}
\usepackage{amsfonts}
\usepackage{mathtools}

\newcommand{\weakid}[1]{\prescript{}{#1}\omega}

\usepackage{amsthm}
\theoremstyle{definition}
\newtheorem{theorem}{Theorem}
\newtheorem*{prf}{Proof}
\newtheorem{lemma}[theorem]{Lemma}
\newtheorem{corollary}[theorem]{Corollary}
\newtheorem{claim}[theorem]{Claim}

\theoremstyle{remark}
\newtheorem{example}{Example}

\begin{document}
	
\title{Final Sentential Forms}

\author{
	Tomáš Kožár
	\institute{Faculty of Information Technology\\
		Brno University of Technology\\
		Czech Republic}
	\email{ikozar@fit.vut.cz}
	\and
	Zbyněk Křivka
	\institute{Faculty of Information Technology\\
		Brno University of Technology\\
		Czech Republic}
	\email{krivka@fit.vut.cz}
	\and
	Alexander Meduna
	\institute{Faculty of Information Technology\\
		Brno University of Technology\\
		Czech Republic}
	\email{meduna@fit.vut.cz}
}
\def\titlerunning{Final Sentential Forms}
\def\authorrunning{A. Meduna, T. Kožár, Z. Křivka}


	\maketitle

	\begin{abstract}
		Let $G$ be a context-free grammar with a total alphabet $V$, and let $F$ be a \textit{final language} over an alphabet $W \subseteq V$. A \textit{final sentential form} is any sentential form of $G$ that, after omitting symbols from $V - W$, it belongs to $F$. The string resulting from the elimination of all nonterminals from $W$ in a final sentential form is in the \textit{language of} $G$ \textit{finalized by} $F$ if and only if it contains only terminals.
		
		The language of any context-free grammar finalized by a regular language is context-free. On the other hand, it is demonstrated that $L$ is a recursively enumerable language if and	only if there exists a propagating context-free grammar $G$ such that $L$ equals the language of $G$ finalized by $\{ w\#w^R \,|\, w \in \{ 0, 1 \}^* \}$, where $w^R$ is the reversal of $w$.
	\end{abstract}

	\section{Introduction}
	The present paper introduces and studies \textit{final sentential forms} of context-free grammars. These forms represent the sentential forms in which the sequences of prescribed symbols, possibly including nonterminals, belong to given \textit{final languages}. If all the other symbols are terminals, these final forms are changed to the sentences of the generated languages by simply eliminating all nonterminals in them. Next, we sketch both a practical inspiration	and a theoretical reason for introducing this new way of context-free language generation. 
	
	\begin{enumerate}
		\item[I.] Indisputably, parsing represents a crucially important application area of ordinary context-free grammars (see Chapters~3 through 5 in~\cite{meduna-compilers}) as well as their modified versions, such as regulated grammars (see Section~20.3 in~\cite{MedZem14}). During the parsing process, the correctness of the source program syntax is often verified before all nonterminals are eliminated; nevertheless, most classically constructed parsers go on  eliminating these nonterminals by using erasing rules until only terminals are derived. As a result, the entire parsing process is slowed down uselessly during this closing phase (for a simple, but straightforward illustration of this computational situation, see, for instance, Case Study~14/35 in~\cite{meduna-compilers} or Example~4.35 in~\cite{Aho}). Clearly, as the newly introduced way of language generation frees us from a necessity of this closing elimination of all nonterminals, the parsers that make use of it work faster.
		
		\item[II.] From a theoretical viewpoint, in the present paper, we achieve a new representation for recursively enumerable languages based upon context-free languages. Admittedly, the theory of formal languages is overflown with many representations for recursively enumerable languages based upon operations over some context-free languages or their special cases (see Section~4.1.3 in~\cite{RozSal97-1}). Nonetheless, we believe this new representation is of some interest when compared with the previously demonstrated representations. Indeed, each of the already existing representations is demonstrated, in essence, by a proof that has the following general format. (i) First, given any recursively enumerable language $L$, it represents $L$ by a suitable language model $G$, such as a phrase structure grammar in a normal form. (ii) Then, from $G$, it derives both operations and context-free languages involved in the representation in question. (iii) Finally, it shows that the representation made in this way from $G$ holds true.  What is important from our standpoint is that in a proof like this, the specific form of all the operations as well as the languages involved in the representation always depend on $G$, which generates $L$. As opposed to this, the new representation achieved in the present paper is much less dependent on $L$ or any of its language models . More precisely, we demonstrate the existence of a unique constant language $C$ defined as $C = \{w\#w^R\,\,|\, w \in \{0, 1\}^*\}$ and express any recursively enumerable language $L$ by using $C$ and a minimal linear language without any operation. Consequently, $C$ always remains unchanged and, therefore, independent of $L$ or its models. Considering this independency as well as the absence of any operations in the new representation, we believe this representation might be of some interest to formal language theory.
	\end{enumerate}
	
	To give a more detailed insight into this study, we first informally recall the notion of an ordinary context-free grammar and its language (this paper assumes a familiarity with formal language theory). A context-free grammar $G$ is based upon a grammatical alphabet $V$ of symbols and a finite set of rules. The alphabet $V$ is divided into two disjoint subalphabets---the alphabet of terminals $T$ and the alphabet of nonterminals $N$. Each rule has the form $A \to x$, where $A$ is a nonterminal and $x$ is a string over $V$. Starting from a special start nonterminal, $G$ repeatedly rewrites strings according to its rules, and in this way, it generates its sentential forms. Sentential forms that consist only of terminal symbols are called sentences, and the set of all sentences represents the language generated by $G$.
	
	In this paper, we shortened the generating process sketched above by introducing a final language $F$ over a subalphabet $W \subseteq V$. A \textit{final sentential form} of $G$ is any of the sentential forms in which the sequence of symbols from $W$ belong to $F$. If in this form, all the symbols from $V-W$ are terminals, the string obtained by eliminating all nonterminals from $N \cap W$ results into a sentence of the generated language $L(G,F)$ finalized by $F$.
	
	\medskip
	
	\noindent Next, we illustrate the newly introduced concept of final sentential forms by a simple example in linguistic morphology, which studies word formation, such as inflection and compounding, in natural languages.

	\medskip
		
	\begin{example}
	Consider an alphabet $\Sigma$ of consonants and vowels. Suppose that a morphological study deals with a language $L$ consisting of all possible words over $\Sigma$ together with their consonant-vowel binary schemes in which every consonant and every vowel are represented by 1 and 0, respectively. Mathematically, $L = \{ w\#\sigma(w) \,|\, w \in \Sigma^+ \}$, where $\sigma$ is the homomorphism from $\Sigma^*$ to $\{0,1\}^*$ defined as $\sigma(x) = 1$ and $\sigma(y) = 0$ for every consonant $x$ in $\Sigma$ and every vowel $y$ in $\Sigma$, respectively. For instance, considering $\Sigma$ as the English alphabet, $the\#110 \in L$ while $the\#100 \not \in L$. Define the context-free grammar $G$ with the following rules.
	
	\begin{itemize}
		\item $S \to A\#B$, $B \to 0YB$, $B \to 0Y$, $B\to 1XB$, $B \to 1X$,
		\item $A \to aAY$, $A \to aY$ for all vowels $a$ in $\Sigma$,
		\item $A \to bAX$, $A \to bX$ for all consonants $b$ in $\Sigma$,
	\end{itemize}
	
	\noindent where the uppercase symbols are nonterminals with $S$ being the start nonterminal, and the other symbols are terminals. Set $W = \{ X,Y,\# \}$ and $F = \{ w\#w^R \,|\, w \in \{X,Y\}^* \}$. For instance, take this step-by-step derivation
	\begin{align*}
		S &\Rightarrow A\#B \Rightarrow tAX\#B \Rightarrow thAXX\#B \Rightarrow theYXX\#B\\
		&\Rightarrow theYXX\#1XB \Rightarrow theYXX\#1X1XB \Rightarrow the YXX\#1X1X0Y
	\end{align*}
	
	\noindent In $theYXX\#1X1X0Y$, $YXX\#XXY \in F$, and apart from $X,Y, \# \in W$, $the YXX\#1X1X0Y$ contains only terminals. The removal of all $X$s and $Y$s in $theYXX\#1X1X0Y$ results into $the\#110$, which thus belongs to $L(G,F)$. 
	On contrary, 
	\begin{align*}
		S &\Rightarrow^* theYXX\#1X1XB \Rightarrow the YXX\#1X1X0YB = \gamma \Rightarrow the YXX\#1X1X0Y0Y = \delta
	\end{align*}
	Let $T = \Sigma \cup \{0,1\}$. Consider $\gamma$. Although $YXX\#XXY \in F$, $the\#110B \notin L(G,F)$ since $B \notin W \cup T$. On the other hand, considering $\delta$, after omitting symbols from $W - T$, we have $the \# 1100 \in T^*$, but since $YXX\#XXYY \notin F$, $the\#1100 \notin L(G,F)$.
	
	Clearly, $L(G,F) = L$.
	\end{example}
	
	As its main result, the present paper demonstrates that $L$ is a recursively enumerable language if and only if $L = L(G,\{ w\#w^R \,|\, w \in \{ 0, 1 \}^* \})$, where $G$ is a context-free grammar; observe that in this equivalence, the final language $\{ w\#w^R \,|\, w \in \{ 0, 1 \}^* \}$ remains constant independently of $L$. On the other hand, the paper also proves that any $L(G,F)$ is context-free if $G$ is a context-free grammar and $F$ is regular.
	
	The rest of the paper is organized as follows. First, Section~\ref{sec:preliminaries} gives all the necessary terminology and defines the new notions, informally sketched in this introduction. Then, Section~\ref{sec:results} establishes the above-mentioned results and points out an open problem related to the present study.
	
	\section{Preliminaries and Definitions}
	\label{sec:preliminaries}
	
	This paper assumes that the reader is familiar with the language theory (see \cite{meduna-FLC}).
	
	For a set $Q$, $card(Q)$ denotes the cardinality of $Q$. For an alphabet $V$, $V^*$ represents the free monoid generated by $V$ under the operation of concatenation. The unit of $V^*$ is denoted by $\varepsilon$. Set $V^+$ = $V^*$ - $\{\varepsilon\}$; algebraically, $V^+$ is thus the free semigroup generated by $V$ under the operation of concatenation. For $w \in V^*$, $|w|$ and $w^R$ denotes the length of $w$ and the reversal of $w$, respectively. Let $W$ be an alphabet and $\omega$  be a homomorphism from $V^*$ to $W^*$ (see \cite{meduna-FLC} for the definition of homomorphism); $\omega$ is a \textit{weak identity} if $\omega(a) \in \{a,\varepsilon\}$ for all $a \in V$.
	
	A \textit{context-free grammar} (CFG for short) is a quadruple $G = (V, T, P, S)$, where $V$ is an alphabet, $T \subseteq V$, $P \subseteq (V-T) \times V^*$ is finite, and $S \in V-T$. Set $N = V - T$. The components $V,T,N,P$, and $S$ are referred to as the total alphabet, the terminal alphabet, the nonterminal alphabet, the set of rules, and the start symbol of $G$, respectively. Instead of $(A ,x) \in P$, we write $A \to x \in P$ throughout. For brevity, we often denote $A \to x$ by a unique label $p$ as $p: A \to x$, and we briefly use $p$ instead of $A \to x$ under this denotation. For every $p: A \to x \in P$, the \textit{left-hand side of} $p$ is defined as $lhs(p) = A$. The grammar $G$ is \textit{propagating} if $A \to x \in P$ implies $x \in V^+$. The grammar $G$ is \textit{linear} if no more than one nonterminal appears on the right-hand side of any rule in $P$. Furthermore, a linear grammar $G$ is \textit{minimal} (see page 76 in \cite{salomaa-formal}) if $N = \{S\}$ and $S \to \# \in P$, $\# \in T$, is the only rule with no nonterminal on the right-hand side, whereas it is assumed that $\#$ does not occur in any other rule. In this paper, a minimal linear grammar $G$ is called a \textit{palindromial grammar} if $card(P) \geq 2$, and every rule of the form $S \to xSy$, where $x,y \in T^*$, satisfies $x=y$ and $x,y \in T$. For instance, $H = (\{ S, 0, 1, \# \}, \{ 0, 1, \# \}, \{ S \to 0S0, S \to 1S1, S \to \# \}, S)$ is a palindromial grammar. 
	
	For every $u, v \in V^*$ and $p: A \to x \in P$, write $uAv \Rightarrow uxv\,[p]$ or, simply, $uAv \Rightarrow uxv$; $\Rightarrow$ is called the \textit{direct derivation} relation over $V^*$. For $n \geq 0, \Rightarrow^n$ denotes the $n$-th power of $\Rightarrow$. Furthermore, $\Rightarrow^+$ and $\Rightarrow^*$ denote the transitive closure and the transitive-reflexive closure of $\Rightarrow$, respectively. Let $\phi(G) = \{w \in V^* |$ $S \Rightarrow^* w\}$ denotes the set of all \textit{sentential forms} of $G$. The language  of $G$ is denoted by $L(G)$ and defined as $L(G) = T^* \cap \, \phi(G)$. For example, $L(H) = \{ w\#w^R \,|\, w \in \{ 0, 1 \}^* \}$, where $H$ is defined as above.
	
	Let $G = (V, T, P, S)$ be a CFG and $W \subseteq V$. Define the weak identity $\weakid{W}$ from $V^*$ to $W^*$ as $\weakid{W}(X) = X$ for all $X \in W$, and $\weakid{W}(X) = \varepsilon$ for all $X \in V - W$. Let $F \subseteq W^*$. Set
	\begin{align*}
		&\phi(G,F) = \{x \,|\, x\in\phi(G),\, \weakid{W}(x) \in F\}\\
		&L(G,F) = \{\, \weakid{T}(y) \,|\, y \in \phi(G,F), \weakid{(N-W)}(y) = \varepsilon \}.
	\end{align*}
	$\phi (G, F)$ and $L(G,F)$ are referred to as the set of \textit{sentential forms of $G$ finalized by $F$} and the \textit{language of $G$ finalized by $F$}, respectively. Members of $\phi(G,F)$ are called \textit{final sentential forms}. $\mathbf{REG}, \mathbf{PAL}, \mathbf{LIN}, \mathbf{CF}$, and $\mathbf{RE}$ denote the families of regular, palindromial, linear, context-free, and recursively enumerable languages, respectively. Observe that 
	
	\[
	\textbf{REG} \cap \textbf{PAL} = \emptyset\text{ and }\mathbf{REG} \cup \mathbf{PAL} \subset \mathbf{LIN}. 
	\]
	
	\noindent Set 
	\begin{align*}
	&\mathbf{CF_{PAL}} = \{L(G,F) \,|\, G \text{ is a CFG, }F \in \mathbf{PAL} \}\\
	&\mathbf{CF_{REG}} = \{L(G,F) \,|\, G \text{ is a CFG, }F \in \mathbf{REG} \}
	\end{align*}
	
	\begin{example}
		Set $I = \{ i(x) \,|\, x\in \{0,1\}^+ \}$, where $i(x)$ denotes the integer represented by $x$ in the standard way; for instance, $i(011) = 3$. Consider 
		\[
		L = \{u\#v \,|\, u,v \in \{0,1\}^+\text{, }i(u) > i(v) \text{ and } |u| = |v|\}.
		\]
		
		\noindent Next, we define a CFG  $G$ and $F \in \mathbf{PAL}$ such that $L = L(G,F)$. Let $G = (V, T, P, S)$ be a context-free grammar. Set $V = \{S,X,\overline{X}, Y, \overline{Y}, A, B, C, D, 0, 1, \#\}$, $T = \{0,1,\#\}$, and set $P$ as the set of the following rules
		
		\begin{itemize}
			\item $S \to X \#\overline{X}$,
			
			\item $X \to 1AX$, $X \to 0BX$, $X \to 1CY$, $X \to 1C$,
			
			\item $\overline{X} \to 1\overline{X}A$, $\overline{X}\to 0\overline{X}B$, $\overline{X}\to 0\overline{Y}C$, $\overline{X}\to 0C$,
			
			\item $Y \to \alpha DY$, $Y\to \alpha D$, $\overline{Y} \to \alpha \overline{Y}D$, $\overline{Y} \to \alpha D$ for all $\alpha \in \{0,1\}$.
		\end{itemize}
	\noindent Set $W = \{A,B,C,D,\#\}$ and $F = \{ w\#w^R \,|\, w \in \{A,B,C,D\}^+ \text{ and } n \geq 1\}$. Observe that $F = L(H)$, where $H = (\{S,A,B,C,D,\#\}, \{A,B,C,D,\#\}, \{S\to ASA, S \to BSB, S\to CSC, S\to DSD, S\to \#\}, S)$ is a palindromial grammar. Therefore, $F \in \mathbf{PAL}$. For instance, take this step-by-step derivation
		\begin{align*}
			S &\Rightarrow X\#\overline{X} \Rightarrow 1AX\#\overline{X} \Rightarrow 1A0BX\#\overline{X} \Rightarrow 1A0B1CY\#\overline{X} \Rightarrow 1A0B1C0D\#\overline{X} \\
			&\Rightarrow 1A0B1C0D\#1\overline{X}A \Rightarrow 1A0B1C0D\#10\overline{X}BA \Rightarrow 1A0B1C0D\#100\overline{Y}CBA  \\
			&\Rightarrow 1A0B1C0D\#1001\overline{Y}DCBA \Rightarrow 1A0B1C0D\#1001DCBA
		\end{align*}
		
	\noindent in $G$. Notice that $\weakid{W} (1A0B1C0D\#1001DCBA) \in F, $ and $\weakid{T}(1A0B1C0D\#1001DCBA) \in L(G,F)$. 
	The reader is encouraged to verify that $L = L(G,F)$.
	\end{example}
	
	A \textit{queue grammar} (see \cite{KleijnRozenberg:AI:1983}) is a sextuple, $Q = (V, T, U, D, s, P)$, where $V$ and $U$ are alphabets satisfying $V \cap\, U = \emptyset$, $T \subseteq V$, $D \subseteq U$, $s \in (V-T)(U-D)$, and $P \subseteq (V\times (U-D)) \times (V^* \times U)$ is a finite relation such that for for every $a \in V$, there exists an element $(a, b, z, c)\in P$. If $u, v \in V^*U$ such that $u = arb; v=rzc;a\in V;r, z \in V^*; b, c \in U;$ and $(a, b, z, c) \in P$, then $u \Rightarrow v$ $[(a, b, z, c)]$ in $G$ or, simply, $u \Rightarrow v$. In the standard manner, extend $\Rightarrow$ to $\Rightarrow^n$, where $n \geq 0;$ then, based on $\Rightarrow^n$, define $\Rightarrow^+$ and $\Rightarrow^*$. The language of $Q$, $L(Q)$, is defined as $L(Q) = \{w \in T^*\,|\, s\Rightarrow^* wf$ where $f \in D\}$.
	A \textit{left-extended queue grammar} is a sextuple, $Q = (V, T, U, D, s, P)$, where $V, T, U, D$, and $s$ have the same meaning as in a queue grammar, and $P\subseteq (V\times (U-D)) \times (V^* \times U)$ is a finite relation (as opposed to an ordinary queue grammar, this definition does not require that for every $a\in V$, there exists an element $(a, b, z, c) \in P)$. Furthermore, assume that $\# \notin V \cup U$. If $u, v \in V^*\{\#\}V^*U$ so that $u=w\#arb$; $v=wa\#rzc$; $a\in V$; $r,z,w \in V^*$; $b, c \in U$; and $(a, b, z, c) \in P$, then $u\Rightarrow v [(a, b, z, c)]$ in $G$ or, simply $u \Rightarrow v$. In the standard manner, extend $\Rightarrow$ to $\Rightarrow^n$, where $n\geq 0;$ then, based on $\Rightarrow^n$, define $\Rightarrow^+$ and $\Rightarrow^*$. The language of $Q$, $L(Q)$, is defined as $L(Q) = \{ v \in T^* \,|\, \#s \Rightarrow^* w\# vf$ for some $w \in V^*$ and $f\in D\}$. Less formally, during every step of a derivation, a left-extended queue grammar shifts the rewritten symbol over $\#;$ in this way, it records the derivation history, which plays a crucial role in the proof of Lemma~\ref{lemma:3} in the next section.
	
	A \textit{deterministic finite automaton} (DFA for short) is a quintuple $M = (Q, \Sigma, R, s, F)$, where $Q$ is a finite \textit{set of states}, $\Sigma$ is an \textit{alphabet of input symbols}, $Q \cap \Sigma = \emptyset$, $s \in Q$ is a special state called the \textit{start state}, $F\subseteq Q$ is a \textit{set of final states} in $M$, and $R$ is a total function from $Q \times \Sigma$ to $Q$. Instead of $R(q,a) = p$, we write $qa \to p$, where $q,p \in Q$ and $a \in \Sigma \cup \{\varepsilon\}$; $R$ is referred to as the \textit{set of rules} in $M$. For any $x \in \Sigma^*$ and $qa\to p \in R$, we write $qax \Rightarrow px$. The \textit{language of} $M$, $L(M)$, is defined as $L(M) = \{ w \,|\, w \in \Sigma^*$, $sw \Rightarrow^* f$, $f \in F \}$, where $\Rightarrow^*$ denotes the reflexive-transitive closure of $\Rightarrow$. Recall that DFAs characterize $\mathbf{REG}$ (see page 29 in \cite{meduna-FLC}).
	
	\section{Results}
	\label{sec:results}
	In this section, we show that every language generated by a context-free grammar finalized by a regular language is context-free (see Theorem~\ref{theorem:regular}). On the other hand, we prove that every recursively enumerable language can be generated by a propagating context-free grammar finalized by 
	$\{w\#w^R \,|\, w\in\{0,1\}^*\}$ (see Theorem~\ref{theorem:RE}).
	
	\begin{lemma}
		\label{lm:context-free}
		Let $G=(V,T,P,S)$ be any CFG and $F \in \mathbf{REG}$. Then, $L(G,F) \in \mathbf{CF}$.
	\end{lemma}
	
	\begin{prf}
		Let $G=(V,T,P,S)$ be any CFG and $F \in \mathbf{REG}$. Let $F = L(M)$, where $M = (Q, W, R, q_s, Q_F)$ is a DFA.
	\end{prf}
	
	\noindent \textit{Construction.}
	Introduce $U = \{ \langle paq \rangle \,|\, p,q \in Q$, $a \in V \} \cup \{ \langle q_sSQ_F\rangle \}$. From $G$ and $M$, construct a new CFG $H$ such that $L(H) = L(G, F)$ in the following way. Set 
	
	\[H = (\overline{V}, T, \overline{P}, \langle q_sSQ_F\rangle )\]\vspace{-6pt}
	
	\noindent The components of $H$ are constructed as follows. Set $\overline{V} = V \cup U$ and initialize $\overline{P}$ to $\emptyset$. Construct $\overline{P}$ as follows:
	
	\begin{enumerate}[label={(\arabic*)}]
		\item[(0)] Add $\langle q_s S Q_F \rangle \to \langle q_sSq_f \rangle$ for all $q_f \in Q_F$.
		\item Let $A \to y_0X_1y_1X_2\dots X_ny_{n} \in P$, where $A \in V-T, y_i \in (V-W)^*$ and $X_j \in V$, $0 \leq i \leq n$, $1 \leq j \leq n$, for some $n \geq 1$; 
		\\then, add $\langle q_1 A q_{n+1}\rangle \to y_0\langle q_1 X_1 q_2 \rangle  y_1 \langle q_2 X_2 q_3 \rangle \dots \langle q_{n} X_n q_{n+1} \rangle y_{n}$ to $\overline{P}$, for all $q_1, q_2, \dots, q_{n+1} \in Q$.
		
		\item Let $A \to \alpha \in P$, where $A \in V-(T \cup W), \alpha \in (V-W)^*$; 
		\\then, add $A \to \alpha$ to $\overline{P}$.
		
		\item Let $\langle p a q \rangle \in U$, where $a \in W \cap T, pa \to q \in R$; 
		\\then, add $\langle p a q \rangle \to a$ to $\overline{P}$.
		
		\item Let $\langle p B q \rangle \in U$, where $pB \to q \in R, B \in W \cap (V-T)$;
		\\then, add $\langle p B q \rangle \to \varepsilon$ to $\overline{P}$.
		
	\end{enumerate}
	To prove $L(G,F) = L(H)$, we first prove $L(H) \subseteq L(G, F)$; then, we establish $L(G, F) \subseteq L(H)$. To demonstrate $L(H) \subseteq L(G,F)$, we first make three observations---(i) through (iii)---concerning every derivation of the form $\langle q_sSq_f\rangle \Rightarrow^* y$ with $y \in T^*$.
	
	\medskip
	
	\noindent (i) By using rules constructed in (1) and (2), $H$ makes a derivation of the form
	
	\[
	\langle q_sSq_f\rangle \Rightarrow^* x_0\langle q_1Z_1q_2\rangle x_1\dots\langle q_{n}Z_nq_{n+1}\rangle x_n
	\]\vspace{-6pt}
	
	\noindent where $x_i \in (T-W)^*$, $0 \leq i \leq n$, $\langle q_j Z_j q_{j+1} \rangle \in U$, $Z_j \in W$, $1 \leq j \leq n$, $q_1 = q_s$, $q_{n+1} = q_f$, $q_1,\dots,q_{n+1} \in Q$, $q_f \in Q_F$.
	
	\medskip
	
	\noindent (ii)
	If 
	
	\[
	\langle q_sSq_f\rangle \Rightarrow^* x_0\langle q_1Z_1q_2\rangle x_1\dots\langle q_{n}Z_nq_{n+1}\rangle x_n
	\]\vspace{-6pt}
	
	\noindent in $H$, then
	
	\[
	S \Rightarrow^* x_0 Z_1 x_1\dots Z_n x_n
	\]\vspace{-6pt}
	
	\noindent in $G$, where all the symbols have the same meaning as in (i).
	
	\medskip
	
	\noindent (iii)
	Let $H$ make
	
	\[
	x_0\langle q_1Z_1q_2\rangle x_1\dots\langle q_{n}Z_nq_{n+1}\rangle x_n \Rightarrow^*y
	\]\vspace{-6pt}
	
	\noindent by using rules constructed in (3) and (4), where $y \in T^*$, and all the other symbols have the same meaning as in (i). Then, for all $1 \leq j \leq n, q_jZ_j \to q_{j+1} \in R, y = x_0U_1x_1\dots U_nx_n$, where $U_j = \weakid{T}(Z_j)$. As $q_jZ_j \to q_{j+1} \in R$, $1 \leq j \leq n$, $q_1=q_s$ and $q_{n+1} = q_f$, $q_f \in Q_F$, we have $Z_1\dots Z_n \in L(M)$.
	
	\medskip
	
	\noindent Based on (i) through (iii), we are now ready to prove $L(H) \subseteq L(G, F)$. Let $y \in L(H)$. Thus, $\langle q_s S Q_F \rangle \Rightarrow^* y$, $y \in T^*$ in $H$. As $H$ is an ordinary CFG, we can always rearrange the applications of rules during $\langle q_s S Q_F \rangle \Rightarrow ^* y$ in such a way that

	\begin{center}
		\begin{tabular}{rllr}
			$\langle q_sSQ_F\rangle$ & $\Rightarrow$ & $\langle q_sSq_f\rangle$ & $(\alpha)$\\
			{} & $\Rightarrow^*$ & $x_0\langle q_1Z_1q_2\rangle x_1\dots\langle q_{m}Z_mq_{m+1}\rangle x_m$ & $(\beta)$\\
			{} & $\Rightarrow^*$ & $y$ & $(\gamma)$\\
		\end{tabular}
	\end{center}
	
	\noindent so that during ($\alpha$), only a rule from (0) is used, during $\beta$ only rules from (1) and (2) are used, and during ($\gamma$) only rules from (3) and (4) are used. Recall that $Z_1Z_2\dots Z_n \in F$ (see (iii)). Consequently, $\weakid{W}(x_0Z_1x_1\dots Z_nx_{n}) \in F$. From (3), (4), (ii), and (iii), it follows that 
	
	\[
	S \Rightarrow^* x_0 Z_1 x_1\dots x_{n-1} Z_n x_n \text{ in } G\textrm{.}
	\]\vspace{-6pt}
	
	\noindent Thus, as $L(M) = F$, we have $y \in L(G, F)$, so $L(H) \subseteq L(G, F)$.
	
	\medskip
	
	\noindent To prove $L(G,F) \subseteq L(H)$, take any $y \in L(G, F)$. Thus, 
	\begin{align*}
		&S \Rightarrow^* x_0Z_1x_1\dots x_{n-1}Z_nx_n \text{ in }G\textrm{, and}\\
		&y =\, \weakid{T}(x_0Z_1x_1\dots x_{n-1}Z_nx_n)\textrm{ with }Z_1\dots Z_n \in F
	\end{align*}
	where 
	$x_i \in (T-W)^*, 0 \leq i \leq n, Z_j \in W, 1 \leq j \leq n$.
	As $Z_1\dots Z_n \in F$, we have $q_1Z_1 \to q_2$, $\dots$, $q_nZ_n \to q_{n+1} \in R$, $q_1,\dots,q_{n+1} \in Q$, $q_1 = q_s$, $q_{n+1} = q_f$, $q_f \in Q_F$. Consequently, from (0) through (4) of the Construction, we see that 
	\begin{align*}
		\langle q_s S Q_f \rangle &\Rightarrow \langle q_s S q_f \rangle\\
		&\Rightarrow^* x_0Z_1x_1\dots Z_nx_n\\
		&\Rightarrow^* x_0U_1x_1\dots U_nx_n
	\end{align*}
	
	\noindent where $U_j =\, \weakid{T}(Z_j)$, $1 \leq j \leq n$. Hence, $y \in L(H)$, so $L(G, F) \subseteq L(H)$.
	
	Thus, $L(G, F) = L(H)$.
	\qed
	
	\begin{theorem}
		\label{theorem:regular}
		$\mathbf{CF_{REG}} = \mathbf{CF}$.
	\end{theorem}
	
	\begin{prf}
		Clearly, $\mathbf{CF} \subseteq \mathbf{CF_{REG}}$. From Lemma~\ref{lm:context-free}, $\mathbf{CF_{REG}} \subseteq \mathbf{CF}$. Thus, Theorem~\ref{theorem:regular} holds true.
		\qed
	\end{prf}
	
	Now, we prove that by using the constant palindromial language $\{w\#w^R\,\,|\, w \in \{0, 1\}^*\}$ to finalize a propagating context-free grammar, we can represent any recursively enumerable language.
	
	\begin{lemma}
		\label{lemma:1}
		Let $L \in \mathbf{RE}$. Then, there exists a left-extended queue grammar $Q$ satisfying $L(Q) = L$.

	\end{lemma}

	\begin{prf}
		See \emph{Lemma} 1 in \cite{meduna-3scg}.
		\qed
	\end{prf}

	\begin{lemma}
	\label{Lemma:2}
		Let $H$ be a left-extended queue grammar.  Then, there exists a left-extended queue grammar, $Q  = (V, T, U, D, s, R)$, such that $L(H) = L(Q)$ and every $(a, b, x, c) \in R$ satisfies $a \in V - T$, $b \in U - D$, $x \in (V - T)^* \cup T^*$, and $c \in U$.  
	\end{lemma}
		
	\begin{prf}
		See \emph{Lemma} 2 in \cite{meduna-3scg}.
		\qed
	\end{prf}
	
	\begin{lemma}
	\label{lemma:3}
		\sloppy Let $Q = (V, T, U, D, s, R)$ be a left-extended queue grammar. Then, $L(Q) = L(G, \{w\#w^R \,|\, w\in\{0,1\}^*\})$, where $G$ is a CFG.
	\end{lemma}

	\begin{prf}
		Without any loss of generality, assume that $Q$ satisfies the properties described in Lemma~\ref{Lemma:2} and that $\{ 0, 1 \} \cap (V \cup U) = \emptyset$. For some positive integer, $n$, define an injection, $\iota$, from $\Psi^*$ to $(\{ 0, 1 \}^n - 1^n)$, where $\Psi = \{ ab \,|\, (a, b, x, c) \in R$, $a \in V-T$, $b \in U-D$, $x \in (V - T)^* \cup T^*$, $c \in U \}$ so that $\iota$ is an injective homomorphism when its domain is extended to $\Psi^*$; after this extension, $\iota$ thus represents an injective homomorphism from $\Psi^*$ to $(\{ 0, 1 \}^n - 1^n)^*$ (a proof that such an injection necessarily exists is simple and left to the reader). Based on $\iota$, define the substitution, $\nu$ from $V$ to $(\{0,1\}^n - 1^n)$ as $\nu (a) = \{\iota(aq) \,|\, q \in U\}$ for every $a \in V$. Extend domain of $\nu$ to $V^*$. Furthermore, define the substitution, $\mu$, from $U$ to $(\{0,1\}^n - 1^n)$ as $\mu(q) = \{\iota(aq)^R \,|\, a \in V\}$ for every $q \in U$. Extend the domain of $\mu$ to $U^*$. Set $J = \{\langle p,i \rangle \,|\, p \in U-D \textrm{ and } i \in \{ 1, 2 \}\}$.
	\end{prf}
	
	\noindent \textit{Construction}.
		Next, we introduce a CFG $G$ so that $L(Q)=L(G,\{w\#w^R \,|\, w\in\{0,1\}^*\})$. Let $G = (\overline{V}, T, P, S)$, where $\overline{V} = J \,\cup\, \{0, 1, \#\} \,\cup\, T$. Construct $P$ in the following way. Initially, set $P = \emptyset$; then, perform the following steps 1 through 5.
		
	\begin{enumerate}
		\item
		if $(a, q, y, p) \in R$, where $a \in V - T$, $p, q \in U - D$, $y \in (V - T)^*$ and $aq = s$,\\
		then add $S \to u\langle p, 1\rangle v$ to $P$, for all $u \in \nu(y)$ and $v \in \mu(p)$;
		
		\item
		if $(a, q, y, p) \in R$, where $a \in V - T$, $p, q \in U - D$ and $y \in (V - T)^*$,\\
		then add $\langle q, 1 \rangle \to u \langle p, 1 \rangle v$ to $P$, for all $u \in \nu(y)$ and $v \in \mu(p)$;
		
		\item
		for every $q \in U - D$, add $\langle q, 1 \rangle \to \langle q, 2 \rangle$ to $P$;
		
		\item
		if $(a, q, y, p) \in R$, where $a \in V - T$, $p, q \in U - D$, $y \in T^*$,\\
		then add $\langle q, 2 \rangle \to y \langle p, 2 \rangle v$ to $P$, for all $v \in \mu(p)$;
		
		\item
		if $(a, q, y, p) \in R$, where $a \in V - T$, $q\in U - D$, $y \in T^*$, and $p \in D$,\\
		then add $\langle q, 2 \rangle \to y\#$ to $P$.
		
	\end{enumerate}

	\noindent Set $W = \{0,1,\#\}$ and $\Omega = \{ xy\#z \in \phi(G) \,|\, x \in \{0,1\}^+$, $y \in T^*$, $z = x^R  \}$.

\begin{claim}
\label{claim:1}
	Every $h \in \Omega$ is generated by $G$ in this way
	
	\smallskip
	
	\noindent
	\begin{tabular}{rl}
		{} & $S$ \\
		$\Rightarrow$ & $g_1\langle q_1, 1 \rangle t_1 \Rightarrow g_2 \langle q_2, 1 \rangle t_2 \Rightarrow \dots \Rightarrow g_{k} \langle q_k, 1 \rangle t_k \Rightarrow g_k \langle q_{k}, 2 \rangle t_{k}$\\
		$\Rightarrow$ & $g_ky_1 \langle q_{k+1}, 2 \rangle t_{k+1} \Rightarrow g_ky_1y_2\langle q_{k+2}, 2 \rangle t_{k+2} \Rightarrow \dots \Rightarrow g_ky_1y_2\dots y_{m-1}\langle q_{k+m-1}, 2 \rangle t_{k + m -1}$\\
		$\Rightarrow$ & $g_ky_1y_2\dots y_{m-1}y_m\#t_{k+m}$
	\end{tabular}
	
	\medskip
	
	\noindent
	in $G$, where $k, m \geq 1$; $q_1,\dots,q_{k+m-1} \in U - D$; $y_1,\dots,y_m \in T^*$; $t_i \in \mu (q_i\dots q_1)$ for $i = 1,\dots,k+m$; $g_j \in \nu(d_1\dots d_j)$ with $d_1,\dots,d_j \in (V - T)^*$ for $j = 1, \dots, k$; $d_1\dots d_k = a_1\dots a_{k+m}$ with $a_1$, $\dots$, $a_{k+m} \in V - T$ (that is, $g_k \in \nu(a_1\dots a_{k+m})$ with $g_k = {(t_{k+m})}^R); h = y_1y_2\dots y_{m-1}y_m$.
	\end{claim}
	
	\begin{prf}

	Examine the construction of $P$. Observe that every derivation begins with an application of a rule having $S$ on its left-hand side. Set $1\textrm{-}J = \{ \langle p, 1 \rangle \,|\, p \in U \}, 2\textrm{-}J = \{ \langle p, 2 \rangle \,|\, p \in U \}, 1\textrm{-}P = \{ p \,|\, p \in P \textrm{ and } lhs(p) \in 1\textrm{-}J \}, 2\textrm{-}P = \{ p \,|\, p \in P \textrm{ and } lhs(p) \in 2\textrm{-}J \}$. Observe that in every successful derivation of $h$, all applications of rules from $1\textrm{-}P$ precede the applications of rules from $2\textrm{-}P$. Thus, the generation of $h$ can be expressed as
	
	\medskip
	
		\noindent\begin{tabular}{rl}
		{} & $S$ \\
		$\Rightarrow$ & $g_1\langle q_1, 1 \rangle t_1 \Rightarrow g_2 \langle q_2, 1 \rangle t_2 \Rightarrow \dots \Rightarrow g_{k} \langle q_k, 1 \rangle t_k \Rightarrow g_k \langle q_{k}, 2 \rangle t_{k}$\\
		$\Rightarrow$ & $g_ky_1 \langle q_{k+1}, 2 \rangle t_{k+1} \Rightarrow g_ky_1y_2\langle q_{k+2}, 2 \rangle t_{k+2} \Rightarrow \dots \Rightarrow g_ky_1y_2\dots y_{m-1}\langle q_{k+m-1}, 2 \rangle t_{k + m -1}$\\
		$\Rightarrow$ & $g_ky_1y_2\dots y_{m-1}y_m\#t_{k+m}$
	\end{tabular}
	
	\medskip
	
	\noindent
	where all the involved symbols have the meaning stated in Claim~\ref{claim:1}.
	\qed
\end{prf}

\begin{claim}
\label{claim:2}
	Every $h \in L(Q)$ is generated by $Q$ in this way
	
	\medskip
	
	\noindent\begin{tabular}{rll}
		{} & $\#a_0q_0$ & {} \\
		
		$\Rightarrow$ & $a_0\#x_0q_1$ & $[(a_0, q_0, z_0, q_1)]$\\
		
		$\Rightarrow$ & $a_0a_1\#x_1q_2$ & $[(a_1, q_1, z_1, q_2)]$\\
		
		\vdots & {} & {}\\
		
		$\Rightarrow$ & $a_0a_1\dots a_k\#x_kq_{k+1}$ & $[(a_k, q_k, z_k, q_{k+1})]$\\
		
		$\Rightarrow$ & $a_0a_1\dots a_ka_{k+1}\#x_{k+1}q_{k+2}$ & $[(a_{k+1}, q_{k+1}, y_1, q_{k+2})]$\\
		
		\vdots & {} & {}\\
		
		$\Rightarrow$ & $a_0a_1\dots a_k a_{k+1}\dots a_{k+m-1}\#x_{k+m-1}y_1\dots y_{m-1}q_{k+m}$ & $[(a_{k+m-1}, q_{k+m-1}, y_{m-1}, q_{k+m})]$\\
		
		$\Rightarrow$ & $a_0a_1\dots a_k a_{k+1}\dots a_{k+m}\#y_1\dots y_mq_{k+m+1}$ & $[(a_{k+m}, q_{k+m}, y_{m}, q_{k+m+1})]$\\
	\end{tabular}
	
	\medskip
	
	\noindent
	where $k, m \geq 1$, $a_i \in V-T$ for $i = 0, \dots, k+m$, $x_j \in (V-T)^*$ for $j = 1, \dots, k+m$, $s = a_0q_0$, $a_jx_j = x_{j-1}z_j$ for $j = 1, \dots, k$, $a_1\dots a_kx_{k+1} = z_0\dots z_k$, $a_{k+1}\dots a_{k+m} = x_k$, $q_0, q_1, \dots, q_{k+m} \in U - D$ and $q_{k+m+1} \in D$, $z_1, \dots, z_k \in (V-T)^*$, $y_1, \dots, y_m \in T^*$, $h = y_1y_2\dots y_{m-1}y_m$.
\end{claim}

\begin{prf}
	Recall that $Q$ satisfies the properties given in Lemma~\ref{Lemma:2}. These properties imply that Claim~\ref{claim:2} holds true.
	\qed
\end{prf}

\begin{claim}
\label{claim:3}
	$L(G,\{w\#w^R \,|\, w\in\{0,1\}^*\}) = L(Q)$.
\end{claim}

\begin{prf}
	To prove that $L(G, \{w\#w^R \,|\, w\in\{0,1\}^*\}) \subseteq L(Q)$, take any $h \in \Omega$ generated in the way described in Claim~\ref{claim:1}. From $\weakid{W}(h) \in \{w\#w^R \,|\, w\in\{0,1\}^*\}$ with $W = \{0, 1, \#\}$, it follows that $xy\#z$ with $z = x^R$ where $x = g_k$, $y = y_1\dots y_m$, $z = t_{k+m}$. At this point, $R$ contains $(a_0, q_0, z_0, q_1)$, $\dots$, $(a_k, q_k, z_k, q_{k+1})$, $(a_{k+1}, q_{k+1}, y_1, q_{k+2})$, $\dots$, $(a_{k+m-1}$, $q_{k+m-1}$, $y_{m-1}$, $q_{k+m})$, $(a_{k+m}$, $q_{k+m}$, $y_m$, $q_{k+m+1})$, where $z_1$, $\dots$, $z_k \in (V - T)^*$, and  $y_1$, $\dots$, $y_m \in T^*$.  Then, $Q$ makes the generation of $\weakid{T}(h)$ in the way described in Claim~\ref{claim:2}. Thus, $\weakid{T}(h) \in L(Q)$.
	
	To prove $L(Q) \subseteq L(G,\{w\#w^R \,|\, w\in\{0,1\}^*\})$, take any $h \in L(Q)$. Recall that $h$ is generated in the way described in Claim~\ref{claim:2}. Consider the rules used in this generation. Furthermore, consider the definition of $\nu$ and $\mu$. Based on this consideration, observe that from the construction of $P$, it follows that $S \Rightarrow^* o h\# \overline{o}$ in $G$ for some $o, \overline{o} \in \{0,1\}^+$ with $\overline{o} = o^R$. Thus, $\weakid{W}(oh\#\overline{o}) \in \{w\#w^R \,|\, w\in\{0,1\}^*\}$, so consequently, $h \in L(G,\{w\#w^R \,|\, w\in\{0,1\}^*\})$.
	\qed
\end{prf}

  \medskip

	\noindent Claims~\ref{claim:1} through \ref{claim:3} imply that Lemma~\ref{lemma:3} holds true.

	\bigskip

	\begin{theorem}
	\label{theorem:RE}
		A language $L \in \mathbf{RE}$ if and only if $L = L(G, \{w\#w^R \,|\, w\in\{0,1\}^*\})$, where $G$ is a propagating CFG.

	\end{theorem}
	
	\begin{prf}
		This theorem follows from Lemmas~\ref{lemma:1} through \ref{lemma:3}.
	\end{prf}
	
	\begin{corollary}
		\label{corollary:re}
		$\mathbf{RE} = \mathbf{CF_{PAL}}$.
	\end{corollary}
	
	\noindent Consider $\{w\#w^R\,|\,w\in \{0,1\}^*\}$ without $\#$---that is $\{ww^R\,|\,w\in \{0,1\}^*\}$. On the one hand, this language is out of $\mathbf{CF_{PAL}}$ because the central symbol $\#$ does not occur in it. On the other hand, it is worth pointing out that Theorem~\ref{theorem:RE} can be based upon this purely binary language as well.
	\begin{corollary}
		A language $L \in \mathbf{RE}$ if and only if $L = L(G, \{ww^R\,|\,w\in \{0,1\}^*\})$, where $G$ is propagating CFG.
	\end{corollary}
	\begin{prf}
		Prove this corollary by analogy with the way Theorem~\ref{theorem:RE} is demonstrated.
	\end{prf}
	
	\noindent Before closing this paper, we point out an open problem. As its main results, the paper has demonstrated that every recursively enumerable language can be generated by a propagating context-free grammar $G$ finalized by $\{ w\#w^R \,|\, w \in \{ 0, 1 \}^* \}$ (see Theorem~\ref{theorem:RE}). Can this results be established with $G$ having a limited number of nonterminals and/or rules?

\section*{Acknowledgement}
This work was supported by Brno University of Technology grant FIT-S-23-8209.

\bibliographystyle{eptcs}
\bibliography{finalsententialforms} 

\end{document}